\documentclass[aip,pof,preprint]{revtex4-1}                  
             
\usepackage{graphicx}
\usepackage{amssymb}
\usepackage{amsmath}
\usepackage{subfigure,wrapfig}

\begin{document}

\title{Four-vortex motion around a circular cylinder}

\author{Marcel N.~Moura }
\author{Giovani L.~Vasconcelos}\email[Corresponding author. Electronic mail: ]{giovani@df.ufpe.br}

\affiliation{Laborat\'orio de F\'{\i}sica Te\'orica e Computacional, 
Departamento de F\'{\i}sica, Universidade Federal de Pernambuco,
50670-901, Recife, Brazil.}

\date{February 13, 2013 }

\begin{abstract}
 The motion of two  pairs of counter-rotating point vortices placed in a  uniform flow past a circular cylinder is studied analytically and numerically.  When the dynamics is restricted to the symmetric subspace---a case that can be realized experimentally by placing a splitter plate in the center  plane---, it is found that there is a family of linearly stable equilibria for same-signed vortex pairs. The nonlinear dynamics  in the symmetric subspace is investigated and several types of orbits are presented.  The analysis reported here provides new insights and reveals  novel features of this four-vortex system, such as the fact that there is no equilibrium for two pairs of vortices of opposite signs on the opposite sides of the cylinder.  (It is argued that such equilibria might exist for vortex flows past a cylinder confined in a channel.) In addition, a new family of  opposite-signed equilibria on the normal line is reported. The stability analysis for antisymmetric perturbations is also carried out and it shows that all equilibria are unstable in this case.
\end{abstract}


\maketitle


\maketitle

\section{Introduction}
\label{intro}

The formation of recirculating eddies in viscous flows past cylindrical structures  is a problem of considerable theoretical interest and practical relevance for many applications  \cite{mmz, sf}. In the well-known case of  flows past a circular cylinder,  a pair of counter-rotating eddies forms behind the cylinder at small Reynolds numbers, which then goes unstable at higher Reynolds numbers and evolves into a von K\'arm\'an vortex street.  This classical problem was first studied by F\"oppl \cite{foeppl} a century ago using a point-vortex model, but only recently it was more fully understood \cite{us}. The motion of multiple vortex pairs in the presence of a cylinder has also attracted considerable attention  \cite{seath,weihs,miller,marsden,shashi2006,borisov2007}.
 Of particular note is the four-vortex configuration recently observed \cite{nature} in the counterflow of superfluid helium II past a circular cylinder, where stationary eddies formed both downstream and upstream of the cylinder. A possible explanation for this unusual vortex arrangement was given in Ref.~[11] in terms of the complex interaction between the normal and superfluid components of He II. It 
 remains an open question whether similar configurations can be observed in classical fluids. Four-vortex motion in an unbounded plane is also of great interest in the context of integrable systems and nonlinear dynamics \cite{eckhardt,aref1,price,rott, aref2, aref3}.

In this paper we investigate the motion of two pairs of point vortices in an inviscid flow past a circular cylinder. First we analyze the dynamics in the symmetric subspace, where the vortices in each pair are symmetrically located with respect to the center plane.  In this setting, we compute symmetric equilibrium configurations  for two identical vortex pairs as well as asymmetric equilibria for nonidentical vortex pairs. We perform the corresponding linear stability analysis, which shows that there is a large subset of these equilibria that are neutrally stable.   The locus of symmetrical equilibria for identical vortex pairs was first found by Elcrat {\it et al.} \cite{elcrat1} but the stability analysis has not been carried out before. On the other hand, the family of asymmetric equilibria for vortex pairs of non-equal strength appears to be new. Since symmetry can be enforced experimentally by attaching splitter plates to the cylinder in the center plane of the flow \cite{foeppl, roshko}, the family of stable equilibria reported here may eventually be of practical relevance.  The nonlinear dynamics in the symmetric subspace is briefly studied numerically and three general classes of orbits are found: i) bounded orbits, ii) semi-bounded orbits, and iii) completely unbounded orbits. As for antisymmetric perturbations, it is shown that the equilibria are always unstable.

We also analyze the  problem of opposite-signed vortex pairs,  in which  case equilibria were known to exist  for  two pairs of vortices  behind the cylinder \cite{seath,weihs}.  Here we present new  equilibrium configurations for the case where the vortices lie on the  {\it normal line}  (i.e., the line bisecting the cylinder perpendicular to the incoming flow). We show furthermore  that there is no equilibrium for two opposite-signed  vortex pairs on the opposite sides of the cylinder, thus correcting an erroneous claim in the literature \cite{shashi2006}.  It is argued, however, that such equilibria are likely to exist for  flows past a cylinder within a rectangular channel, which might help to explain the unusual vortex configuration seen in superfluid helium mentioned above. A more detailed study of the interesting but more difficult problem of vortex flows past a cylinder in confined geometries is beyond the scope of the present paper.
  
The paper is organized as follows. In Sec.~\ref{sec:2} we present the mathematical formulation of the problem. In Sec.~\ref{sec:3} we study the dynamics of our four-vortex system in the symmetric subspace. In particular, we compute equilibrium configurations for same-signed vortex pairs and study their stability properties. Opposite-signed equilibria on the normal line are also presented, and the nonlinear dynamics for identical pairs of vortices is discussed. The linear stability analysis  for anti-symmetric perturbations is presented in Sec.~\ref{sec:4} and some important implications of our results are discussed in Sec.~\ref{sec:5}. In Sec.~\ref{sec:6} we summarize our main findings and conclusions.

\section{Problem Formulation}
\label{sec:2}

\begin{figure}[t]
\includegraphics[width=0.45\textwidth]{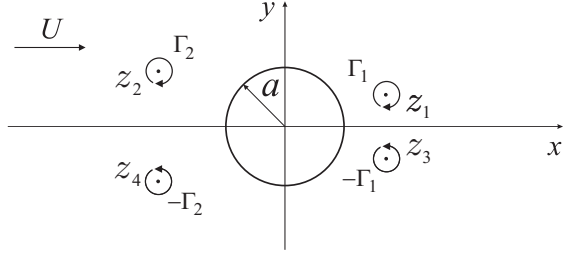}
\caption{Two vortex pairs in a flow past a circular cylinder.}
\label{fig:1}
\end{figure}

We consider the two-dimensional motion of two pairs of vortices around a circular cylinder of radius $a$, in the presence of a uniform stream of velocity $U$, as illustrated in Fig.~\ref{fig:1}. The vortices are considered to be point-like and the fluid is treated as incompressible, inviscid, and irrotational, except at the vortex positions where the vorticity is singular (i.e., a delta function). Under such conditions one has a  potential  flow: the fluid velocity field is given by $\vec{v}=\vec{\nabla}\phi$, where $\phi(x,y)$ is the velocity potential which satisfies Laplace equation, $\nabla^2\phi=0$. 
It is convenient to work in the complex $z$-plane, where $z=x+iy$, with the origin placed at the center of the cylinder. The upper and lower vortices of the vortex pair  downstream of the cylinder are located at positions $z_1=x_1 + iy_1$ and $z_3=x_3 + iy_3$, and have circulations $\pm\Gamma_1$, respectively, whereas the positions of the upper and lower vortices of the vortex pair upstream of the cylinder are denoted by $z_2=x_2 + iy_2$ and $z_4=x_4 + iy_4$, with respective circulations denoted by $\pm \Gamma_2$; see Fig.~\ref{fig:1}.

The complex potential
$w(z)=\phi(x,y)+i\psi(x,y)$ for the flow, where  $\psi$ is the stream function,  is obtained by a direct application of the circle theorem \cite{milne}, yielding\hspace{10cm}
\begin{align}
w(z)=U\left(z+\frac{a^2}{z}\right)+\frac{\Gamma_1}{2 \pi i} \log\frac{\left(z-z_1\right)\left(z-a^2/\bar{z}_3\right)}{\left(z-a^2/\bar{z}_1\right)\left(z-z_3\right)} +
 \frac{\Gamma_2}{2 \pi i} \log\frac{\left(z-z_2\right)\left(z-a^2/\bar{z}_4\right)}{\left(z-a^2/\bar{z}_2\right)\left(z-z_4\right)} ,
\label{w4v}
\end{align}
where the bar denotes complex conjugation. In the right-hand side of Eq.~(\ref{w4v}), the first two terms account for the uniform stream and its image by the cylinder (a dipole at the origin), whereas the other two terms represent the contributions from  the vortices at $z_j$, $j=1,2,3,4$,  and their respective images which are located (inside the cylinder) at $a^2/\bar{z}_j$. 

Introducing dimensionless variables
\begin{equation}
z'=\frac{z}{a}, \quad t'=\frac{U}{a}t, \quad w'=\frac{w}{Ua}, \quad \kappa_i=-\frac{\Gamma_i}{2\pi Ua} , 
\label{dim4v}
\end{equation}
Eq.~(\ref{w4v}) can be rewritten as
\begin{equation}
w(z)=z+\frac{1}{z}+i \kappa_1 \log\frac{\left(z-z_1\right)\left(1-\bar{z}_3 z\right)}{\left(1-\bar{z}_1 z\right)\left(z-z_3\right)} 
+i \kappa_2 \log\frac{\left(z-z_2\right)\left(1-\bar{z}_4 z\right)}{\left(1-\bar{z}_2 z\right)\left(z-z_4\right)}  ,
\label{wdim4v}
\end{equation}
where the prime notation has been dropped. 
To calculate the velocity, $\vec{v}_j=(u_j,v_j)$, of a given  vortex located  at position  $z_j$, one must subtract from the complex potential (\ref{wdim4v}) the contribution of the vortex itself and then evaluate the derivative of the resulting ``effective potential'' at the vortex position $z=z_j$. For example, for the upper vortex located at  $z_1$ one has  
\begin{eqnarray}
u_{1}-iv_{1}&=&\frac{d}{dz}\left.\bigg[ w(z)-i\kappa_1 \log
(z-{z_1})\bigg]\right|_{z=z_1} ,
\label{eq:22}
\end{eqnarray}
which yields
\begin{align}
u_1-iv_1=&~1-\frac{1}{z_1^2}+i \kappa_1  \left(-\frac{1}{z_1-z_3}-\frac{\bar{z}_3}{1-z_1\bar{z}_3} +  \frac{\bar{z}_1}{1-z_1\bar{z}_1}\right)\cr
&+ i \kappa_2  \left(-\frac{1}{z_1-z_4}
 + \frac{1}{z_1-z_2}+ \frac{\bar{z}_2}{1-z_1 \bar{z}_2}-\frac{\bar{z}_4}{1-z_1\bar{z}_4}\right) .
\label{eq:u1}
\end{align}
Similar procedure gives the velocity $\vec{v}_2=(u_2,v_2)$ for the second upper vortex  at $z_2$:
\begin{align}
u_2-iv_2 =&~1-\frac{1}{z_2^2}+i \kappa_2  \left(-\frac{1}{z_2-z_4}-\frac{\bar{z}_4}{1-z_2\bar{z}_4} +  \frac{\bar{z}_2}{1-z_2\bar{z}_2}\right)\cr
&+ i \kappa_1  \left(-\frac{1}{z_2-z_3}
 + \frac{1}{z_2-z_1}+ \frac{\bar{z}_1}{1-z_2 \bar{z}_1}-\frac{\bar{z}_3}{1-z_2\bar{z}_3}\right) .
\label{eq:u2}
\end{align}
The velocity  of the lower vortices can be obtained from  Eq.~(\ref{eq:u1})  by a proper interchange of the indexes: for the vortex located at $z_3$ one makes $1\leftrightarrow3$ and $2\leftrightarrow4$, whereas for  the vortex at $z_4$ one takes  $1\leftrightarrow4$  and $2\leftrightarrow3$, together with the change $\kappa_j\rightarrow-\kappa_j$.

As is well known, the equations of motion for point vortices in a two-dimensional inviscid flow can be formulated as a Hamiltonian system \cite{saffman}.  The dynamics of point vortices in the presence of  rigid boundaries  was shown by Lin \cite{lin1941} to be also Hamiltonian with the same canonical symplectic structure as in the absence of boundaries.   For the problem of two pairs of vortices around a circular cylinder  the corresponding phase space is eight-dimensional, and the Hamiltonian can be obtained explicitly \cite{shashi2006}  but this is not necessary for our purposes.  
Here we are primarily interested in finding the equilibrium positions of this vortex system and studying their linear stability properties. Some interesting aspects of the nonlinear dynamics that ensues when the respective equilibria are perturbed will also be discussed. We start our analysis by considering the dynamics in the four-dimensional symmetric subspace, where the upper and lower vortices in each vortex pair are located at symmetrical positions with respect to the $x$ axis.  The nonsymmetric dynamics will be be discussed afterwards.

\section{Dynamics on the Symmetric Subspace}
\label{sec:3}

It is not difficult to see from Eqs.~(\ref{eq:u1}) and (\ref{eq:u2}) [and the corresponding equations for vortices 3 and 4] that if the vortices are initially placed at positions symmetrically located  with respect to the
  centerline, i.e., $z_3(0)= \overline{z}_{1}(0)$ and $z_4(0)= \overline{z}_{2}(0)$, then this symmetry is preserved for all times.  Because of this symmetry, in this section we shall fix our attention only on the two upper vortices, with the understanding that the location of the lower vortices will correspond to the mirror
  images (with respect to the centerline) of the respective upper vortices. As already noted,  symmetry can be enforced experimentally by placing splitter plates in front and behind the cylinder in the center plane of the flow \cite{foeppl,roshko}, and so the results of this section may be of practical relevance for real flows, as will be discussed later.

With $z_3=\bar{z}_1$ and $z_4=\bar{z}_2$, Eqs.~(\ref{eq:u1}) and (\ref{eq:u2})  can be more conveniently expressed as
\begin{align}
u_1-iv_1=&~1-\frac{1}{z_1^2}+2 \kappa_1 y_1 \left[\frac{1}{(1-r_1^2)(1-z_1^2)}-\frac{1}{4y_1^2} \right]\cr
&+ 2 \kappa_2  y_2 \left[\frac{1}{1-2x_2z_1+r_2^2z_1^2}-\frac{1}{r_2^2-2x_2z_1+z_1^2} \right] 
\label{eq:u1s}
\end{align}
and
\begin{align}
u_2-iv_2 =&~1-\frac{1}{z_2^2}+2 \kappa_2 y_2 \left[\frac{1}{(1-r_2^2)(1-z_2^2)}-\frac{1}{4y_2^2} \right]\cr
&+ 2 \kappa_1  y_1 \left[\frac{1}{1-2x_1z_2+r_1^2z_2^2}-\frac{1}{r_1^2-2x_1z_2+z_2^2} \right],
\label{eq:u2s}
\end{align}
where $r_i^2=x_i^2+y_i^2$, for $i=1,2$.

\subsection{Equilibrium Configurations}
\label{sec:3aa}

The equilibrium positions for the vortex system above are obtained by solving Eqs.~(\ref{eq:u1s}) and (\ref{eq:u2s}) for $u_j=v_j=0$, $j=1,2$. For $\kappa_1\ne\kappa_2$ this amounts to finding the zeros of polynomials of very high order. The problem is relatively easier when the two vortex pairs have the same strength, i.e., $|\kappa_1|=|\kappa_2|$,  as discussed next.

\subsubsection{Same-Signed Equilibria}
\label{sec:3a}

We assume here that the two vortex pairs have the same sign,  i.e., $\kappa_1\kappa_2>0$. Let us consider first the case of equal strength, $\kappa_1=\kappa_2=\kappa$. From symmetry considerations, it is clear that the equilibrium configuration in this case must be such that the vortices are located at  
\begin{align}
z_1=z_0 \quad\mbox{and}\quad  z_2=-\bar{z}_0.
\label{eq:s}
\end{align}
A necessary condition
for a stationary configuration to exist is that the upper (lower) vortices be
of negative (positive) circulation, hence only the case $\kappa>0$ is of
interest here; see Eq.~(\ref{dim4v}).
In this case, it is easy to convince oneself that if we happen to find a configuration in which the velocity $(u_1,v_1)$ of the first vortex  vanishes, then the velocity $\vec{v}_2$ of the second vortex will also vanish. The problem thus reduces to solving  Eq.~(\ref{eq:u1s}) for $u_1=v_1=0$, with  $\kappa_1=\kappa_2$ and the condition (\ref{eq:s}).

\begin{figure}[t]
\subfigure[\label{fig:s}]{\includegraphics[width=0.45\textwidth]{fig2.eps}}\qquad
\subfigure[\label{fig:k}]{\includegraphics[width=0.45\textwidth]{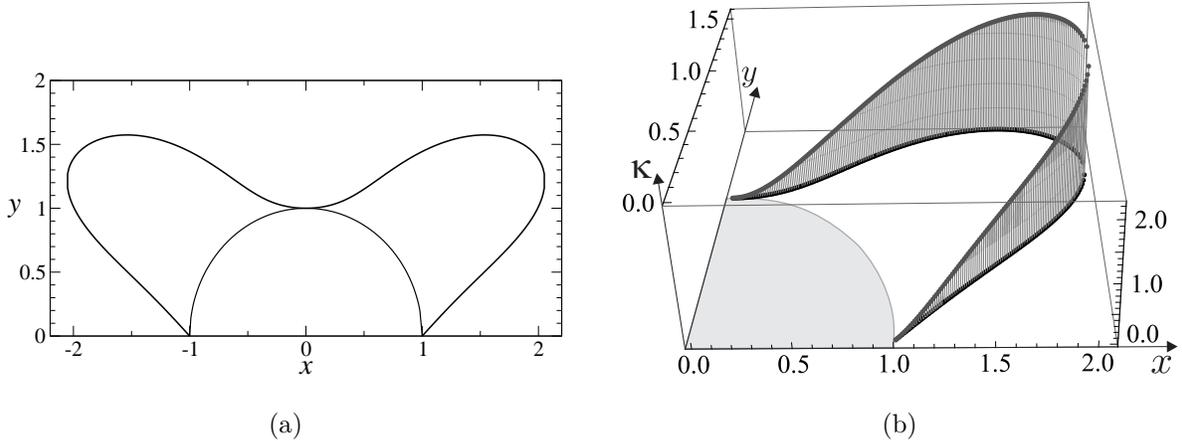}}
\caption{(a) Locus of symmetric equilibria for two identical vortex pairs; only the locations of the upper vortices are shown. (b) Vortex intensity $\kappa$ along the curve $C_0$ of symmetric  equilibria.}
\label{fig:2}
\end{figure}

After  setting $x_1=-x_2=x$, $y_1=y_2=y$, and  $\kappa_1=\kappa_2=\kappa$ in Eq.~(\ref{eq:u1s}) and performing some algebraic manipulation, one finds that the locus of possible equilibrium positions for the first vortex is obtained by solving the equation $P(x,y)=0$, where  $P(x,y)$ is a polynomial of order 14 given in Eq.~(\ref{eq:P2}) of Appendix \ref{app:A}.  Solving this equation in the first quadrant yields the curve $C_0$  shown in Fig.~\ref{fig:s}, with the equilibrium positions for the second vortex being obtained by a reflection of $C_0$ about the $y$ axes. For each point $(x,y)$ on the curve $C_0$, the corresponding vortex intensity $\kappa$ is given by
\begin{equation}
\kappa=\frac{Q(x,y)}{R(x,y)},
\end{equation}
where $Q(x,y)$ and $R(x,y)$ are polynomials given in Eqs.~(\ref{eq:Q}) and (\ref{eq:R}), respectively.
 Fig.~\ref{fig:k} shows a plot of  the vortex intensity $\kappa$ for points on the curve $C_0$.

It is interesting to note that, differently from the case of a single vortex pair behind a cylinder \cite{us}, the stationary positions in the four-vortex case lie in a bounded region close to the cylinder. In other words, equilibrium configurations exist only up to a certain maximum vortex strength [see  Fig.~\ref{fig:k}], beyond which the vortex-vortex interactions cannot be cancelled by the oncoming stream. Note also from Fig.~\ref{fig:k} that for each value of $\kappa$ (in the allowed range) there are two possible equilibria: one closer behind the cylinder and the second one closer to the cylinder top. The symmetric equilibria shown in Fig.~\ref{fig:s} were first found numerically by Elcrat et al.~\cite{elcrat1}. They were also obtained  by Shashikanth \cite{shashi2006} who considered the problem of two symmetric pairs of point vortices interacting with a neutrally buoyant  cylinder, but there the stability properties of the equilibria are quite different from the  case of a fixed cylinder studied here; see below.

\begin{figure}[t]
\includegraphics[width=0.6\textwidth]{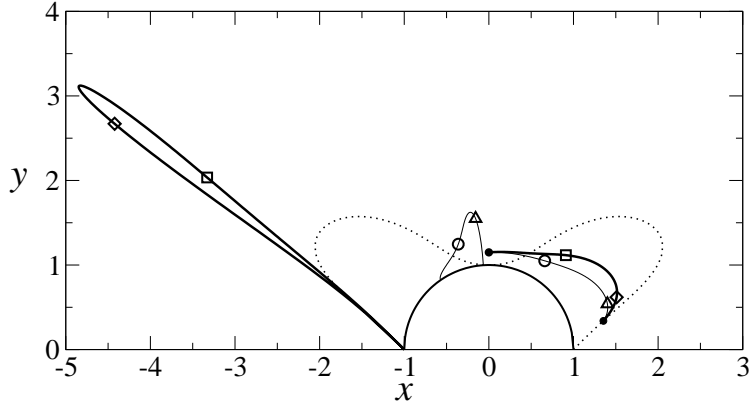}
\caption{The two families of equilibria  (thick and thin solid lines) for two same-signed vortex pairs of different strength, with $\kappa_1=0.5$. In each family of solutions, two particular equilibrium configurations are indicated, for which $\kappa_2=5.356$ (diamonds), $\kappa_2=3.963$ (squares), $\kappa_2=1.414$ (triangles), and $\kappa_2=0.774$ (circles). The black dots indicate the two equilibrium points for a single vortex pair of strength $\kappa=0.5$, and the dotted line represents the symmetrical equilibria for two identical vortex pairs shown in Fig.~\ref{fig:2}.}
\label{fig:dk}
\end{figure}

\begin{figure}[t]
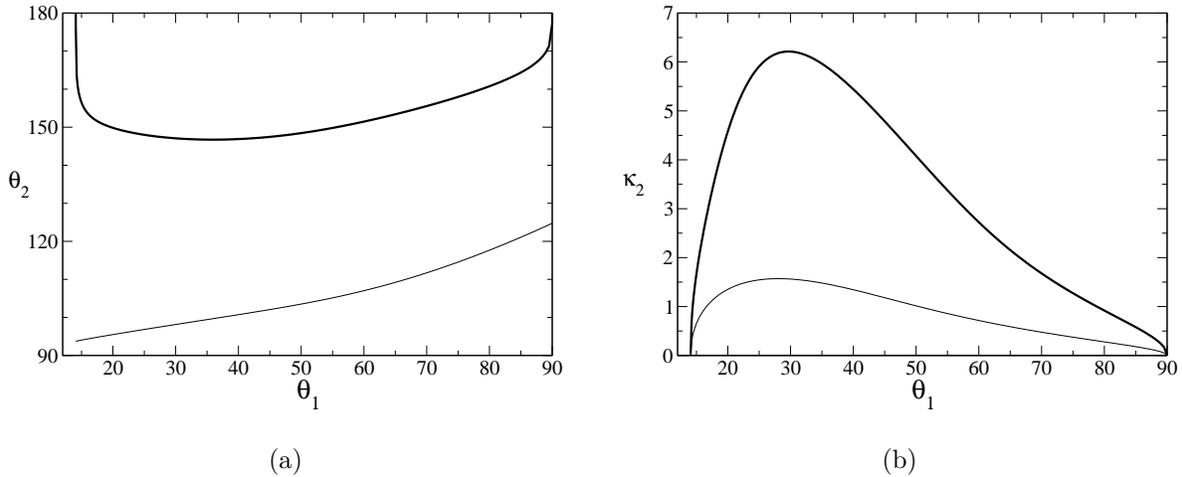

\vspace{0.5cm}
\subfigure[\label{fig:theta2}]{\includegraphics[width=0.45\textwidth]{fig5_new.eps}}\qquad
\subfigure[\label{fig:k2}]{\includegraphics[width=0.45\textwidth]{fig6_new.eps}}
\caption{ Angular position (a) and strength (b) of the second vortex as a function of the angular position of the first vortex for the two families of equilibria  shown in Fig.~\ref{fig:dk}.}
\end{figure}

The symmetry condition (\ref{eq:s}) can be relaxed if one allows for different vortex strengths, i.e., $\kappa_1\ne\kappa_2$.  The set of equilibria in this case can be computed by  varying the parameters $\kappa_1$ and $\kappa_2$, so that for each pair of values $(\kappa_1,\kappa_2)$ one needs to solve the equilibrium equations numerically to obtain the vortex locations $z_1$ and $z_2$.  One convenient way to compute the equilibria  in this case is to fix the  strength of one of the vortex pairs, say, $\kappa_1$, and then vary the strength of the other.  In this case, one typically finds two families of equilibria, which extend the two equilibrium points in the curve $C_0$. In Fig.~\ref{fig:dk}  we show the loci (thick and thin solid lines) of the two families of asymmetric equilibria for $\kappa_1=0.5$. Two particular equilibrium configurations for each family are indicated in Fig.~\ref{fig:dk}, corresponding to $\kappa_2=5.356$ (diamonds), $\kappa_2=3.963$ (squares), $\kappa_2=1.414$ (triangles), and $\kappa_2=0.774$ (circles).  Note that the endpoints of the locus of equilibria of the first vortex in both families (thick and thin solid lines in the first quadrant) are the two corresponding equilibria of a single vortex pair, namely,  the   F\"oppl equilibrium (lower black dot)  and the  equilibrium on the normal line (upper black dot), as expected, since  $\kappa_2=0$ in both points. 
In  Fig.~\ref{fig:theta2} we plot the polar angle,   $\theta_2$ , of the second vortex as a function of the corresponding  angular position, $\theta_1$, of the first vortex,
while in  Fig.~\ref{fig:k2} we plot the corresponding values of $\kappa_2$ as a function of  $\theta_1$.
Notice that in both curves in Fig.~\ref{fig:k2},  for each value of $\kappa_2$, there are two possible equilibria, in analogy to what is observed for the symmetric equilibria; compare with Fig.~\ref{fig:k}.

\subsubsection{Equilibria on the normal line}

\label{sec:normal}

\begin{figure}[t]
\includegraphics[width=0.5\textwidth]{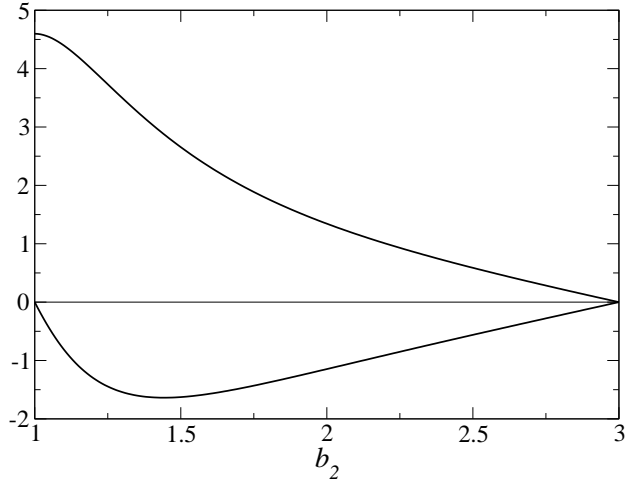}
\caption{Vortex strength $\kappa_1$ (upper curve) and $\kappa_2$ (lower curve) as a function of the location $b_2$ of the second vortex for the equilibrium point on the normal line with $b_1=3$.}
\label{fig:normal}
\end{figure}

Equilibrium configurations   have long been known  to exist  \cite{seath, weihs} for two opposite-signed  vortex pairs  (of unequal strength)  behind the cylinder. Here we present a new family of equilibria in which the vortices  are located on the line bisecting the cylinder perpendicularly to the incoming flow. In this case, the equilibrium positions of the upper vortices are given by
\begin{equation}
x_{1}=0, \quad y_{1}=b_{1} \qquad  \mbox{and}\qquad x_{2}=0, \quad y_{2}= b_{2},
\label{eq:b1}
\end{equation} 
where without loss of generality we take $1<b_{2}<b_{1}$. 
Setting $z_1=ib_1$ and $z_2=ib_2$ in Eqs.~(\ref{eq:u1s}) and (\ref{eq:u2s}),  one immediately sees that the respective right-hand sides are purely real, hence $v_1=v_2=0$, as required by symmetry considerations.  Then equating $u_1=u_{2}=0$  and performing some simplification, one obtains the following system of linear algebraic equations that determine  $\kappa_1$ and $\kappa_2$:
\begin{align}
{\kappa_1} -   \frac{2  b_1^2b_2(b_2^2-1) }{(b_1^2 b_2^2-1)(b_1^2-b_2^2)} \, \kappa_{n}^{(1)} \kappa_2 = \kappa_{n}^{(1)} \, ,
\label{eq:k13}
\end{align}
\begin{align}
{\kappa_2} +  \frac{2  b_2^2b_1(b_1^2-1) }{(b_1^2 b_2^2-1)(b_1^2-b_2^2)} \, \kappa_{n}^{(2)} \kappa_1 &= \kappa_{n}^{(2)} \, .\
\label{eq:k23}
\end{align}
where
\begin{align}
\kappa_{n}^{(i)} =\frac{2(b_i^2-1)(b_i^2+1)^2}{b_i(b_i^4+4b_i^2-1)}, \quad\mbox{for}\quad i=1,2,
\end{align}
is the vortex strength for the corresponding equilibrium  $(z=ib_i)$  on the normal line for a single pair of vortices \cite{us}. Solving Eqs.~(\ref{eq:k13}) and (\ref{eq:k23}) yields
\begin{align}
\kappa_{1}= \frac{\kappa_{n}^{(1)}\left(1+C_{1}\kappa_{n}^{(2)}\right)}{1+C_{1}C_{2}\kappa_{n}^{(1)}\kappa_{n}^{(2)}} ,
\end{align}
\begin{align}
\kappa_{2}  = \frac{\kappa_{n}^{(2)}\left(1-C_{2}\kappa_{n}^{(1)}\right)}{1+C_{1}C_{2}\kappa_{n}^{(1)}\kappa_{n}^{(2)}},
\end{align}
where
\begin{align}
C_{i} = \frac{2b_{1}b_{2}(b_{1}^{2}-1)(b_{2}^{2}-1)}{(b_{1}^{2}b_{2}^{2}-1)(b_{1}^{2}-b_{2}^{2})}\left(\frac{b_i}{b_{i}^{2}-1}\right).
\end{align} 
In Fig.~\ref{fig:normal} we plot  $\kappa_1$ and $\kappa_2$ as a function of $b_2$ for the case $b_1=3$. One sees from this figure that the vortex farthest away from the cylinder has  a negative circulation (since $\kappa_1>0$), whereas the vortex closest  to the cylinder has a positive circulation (i.e., $\kappa_2<0$).

Equilibrium configurations for two opposite-signed vortex pairs do not seem to exist when the vortex pairs are on the opposite sides of the cylinder, i.e., one vortex pair in front of the cylinder and the second vortex pair (of opposite polarity) behind it.
For example, one can easily prove that there are no such equilibria  for vortex pairs of equal strength. (In Ref.~[9]  it was erroneously claimed that such configurations exist.) To see this, first note that  for opposite-signed  vortex pairs of equal strength the fact that the velocity of the first pair vanishes does not automatically ensure that the velocity of the second pair also vanishes. Indeed, setting $\kappa_1=-\kappa_2=\kappa$  in Eq.~(\ref{eq:u1}) and solving for $u_1=v_1=0$, under condition (\ref{eq:s}), yields  a polynomial curve of the form $P_1(x_0,y_0)=0$ for the putative equilibrium (this is the curve shown in Fig.~15 of Ref.~[9]), with a vortex strength given by a rational function: $\kappa=Q_1(x_0,y_0)/R_1(x_0,y_0)>0$. 
%[The polynomials $P_1(x,y)$, $Q_1(x,y)$, and $R_1(x,y)$ are not given here for brevity.] 
However, when solving Eq.~(\ref{eq:u2}) for $u_2=v_2=0$ one finds a vortex strength of the form $\kappa=-Q_1(x_0,y_0)/R_1(x_0,y_0)<0$, in contradiction with the previous result. Hence, no symmetric equilibrium  is possible for opposite-signed vortex pairs of equal strength. For the case of two opposite-signed vortex pairs of unequal strength (on the opposite sides of the cylinder), we have performed a  numerical search for  equilibria by solving the appropriate polynomial equations with $|\kappa_1|\ne|\kappa_2|$ and failed to obtain any valid solution. We conjecture, however, that  if the cylinder is confined within a  rectangular channel such equilibria should appear, owing to  the presence of the channel walls and the infinitely many vortex images that they entail; see Sec.~\ref{sec:5} for further discussion about this problem.

\subsubsection{Equilibrium at infinity}
\label{sec:infty}

Equations (\ref{eq:u1s})  and (\ref{eq:u2s}) also  admit an  equilibrium point at infinity for which the positions of the two vortices are given by
\begin{equation}
x_{1}=\infty, \quad y_{1}=\frac{\kappa_{1}}{2}  \qquad \mbox{and}\qquad x_{2}=-\infty, \quad y_{2}=\frac{\kappa_{2}}{2} .
\label{eq:yc}
\end{equation}
The physical origin of this equilibrium can be easily understood \cite{us}: since the two vortex pairs are infinitely separated from one another, the interaction between them  becomes
negligible and so a stationary configuration is possible if the vortices 
in each pair are separated
by the appropriate distance ($d=\kappa_{i}$), such that the velocity induced by  one of the vortices on the other vortex precisely cancels out the velocity of the oncoming stream.
Notice that this equilibrium is analogous to the two equilibrium points  that exist  at infinity  (i.e., at $x=\pm\infty$, $y=\kappa/2$) for a single pair of vortices  \cite{us}, with the difference that, here, there is one vortex pair in each of these equilibrium points.

\subsection{Linear Stability Analysis}
\label{sec:4a}

Let us denote by $z_{10}$ and $z_{20}$ a generic equilibrium of the vortex system described above and consider arbitrary perturbations (in the symmetric subspace) of the form
\begin{align}
 z_1=z_{10}+\xi_1+i\eta_1, \qquad  z_2=z_{20}+\xi_2+i\eta_2,
 \label{eq:delta}
 \end{align}
where $\xi_i$ and $\eta_i$ are (infinitesimally small) real numbers. 
After inserting Eqs.~(\ref{eq:delta})  into Eqs.~(\ref{eq:u1}) and (\ref{eq:u2}) and linearizing   the resulting equations of motion for $\xi_i$ and $\eta_i$, one obtains the following dynamical system:
\begin{equation}
 \left(\begin{array}{c} \dot{\xi_1}\cr \dot{\eta_1}\cr \dot{\xi_2}\cr \dot{\eta_2}\end{array}\right)= 
A  \left(\begin{array}{c}{\xi_1}\cr{\eta_1}\cr{\xi_2}\cr{\eta_2}
\end{array}\right),
\end{equation}
where $A$ is  a $4\times4$ matrix given by
\begin{equation}
 A
=\left(
\begin{array}{cccc}
 \frac{\partial u_1}{\partial x_1} & \frac{\partial u_1}{\partial y_1} & \frac{\partial u_1}{\partial x_2} & \frac{\partial u_1}{\partial y_2} \\ \\
 \frac{\partial v_1}{\partial x_1} & \frac{\partial v_1}{\partial y_1} & \frac{\partial v_1}{\partial x_2} & \frac{\partial v_1}{\partial y_2} \\ \\
 \frac{\partial u_2}{\partial x_1} & \frac{\partial u_2}{\partial y_1} & \frac{\partial u_2}{\partial x_2} & \frac{\partial u_2}{\partial y_2} \\ \\
 \frac{\partial v_2}{\partial x_1} & \frac{\partial v_2}{\partial y_1} & \frac{\partial v_2}{\partial x_2} & \frac{\partial v_2}{\partial y_2}
\end{array}
\right),
\label{eq:A}
\end{equation}
with the derivatives evaluated at the equilibrium position $z_{10}$ and $z_{20}$.  

Since in the symmetric subspace we have a 4D Hamiltonian system, it follows that the eigenvalues of the matrix $A$ are  of the form: $\lambda_1^\pm=\pm\sqrt{\lambda_1^2}$ and $\lambda_2^\pm=\pm\sqrt{\lambda_2^2}$. The stability of the respective equilibrium will thus  be determined by the sign of  the squared eigenvalues  $\lambda_1^2$ and $\lambda_2^2$. If both quantities are negative then the eigenvalues are purely imaginary, in which case the equilibrium is neutrally stable. On the other hand, if either $\lambda_1^2$ or $\lambda_2^2$ is positive then there is at least one positive real eigenvalue and the equilibrium is therefore unstable. Next we shall carry out the linear stability analysis for the equilibria reported in the preceding subsection.

\subsubsection{Same-Signed Equilibria}
\label{sec:4aa}

\begin{figure}[t]
\includegraphics[width=0.5\textwidth]{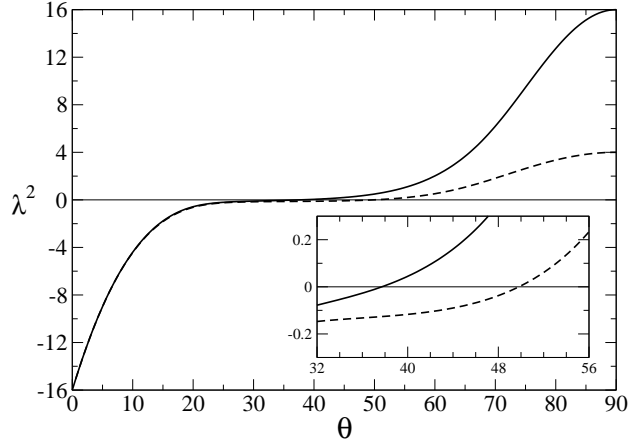}
\caption{Squared eigenvalues $\lambda_1^2$ (solid lines) and $\lambda_2^2$ (dashed lines) for symmetric perturbations of the symmetric equilibria as a function of the angle $\theta$ on the curve $C_0$.}
\label{fig:l2s}
\end{figure}

In the case of two same-signed vortex pairs of equal strength the equilibria are given by $z_{10}=z_0$ and $z_{20}=-\bar{z}_0$, where $z_0=x+iy$ lies on the curve $C_0$ shown in  Fig.~\ref{fig:s}. In this case it turns out  that only six elements of the matrix $A$ given in Eq.~(\ref{eq:A}) are independent of one another. These elements can be written explicitly  in terms of rational functions of the coordinates $(x,y)$ and are given in Appendix \ref{app:B}.

From the  analysis of the eigenvalues of  $A$ one finds that
the stability nature of the equilibrium point varies with its location along the curve $C_0$.   This is indicated in Fig.~\ref{fig:l2s} where we plot the squared eigenvalues $\lambda_1^2$ and $\lambda_2^2$ as function of the polar angle $\theta$ along the curve $C_0$. One sees from this figure that both $\lambda_1^2$ and $\lambda_2^2$ are negative for small angles. Then, as $\theta$ increases, $\lambda_1^2$  vanishes around $\theta=38^\circ$ and becomes positive for the remaining points, with similar change of sign occurring for $\lambda_2^2$  around $\theta=50^\circ$; see inset of Fig.~\ref{fig:l2s}. The equilibrium is thus a center-center \cite{hs} (in the symmetric subspace) for $0^\circ<\theta <38^\circ$, becomes a saddle-center  for $38^\circ<\theta <50^\circ$, and turns into a saddle-saddle  for $50^\circ<\theta <90^\circ$.  
We note furthermore that the maximum of $\kappa$ in Fig.~\ref{fig:k} takes place at  about $\theta=38^\circ$, and so the first of the two possible equilibria for a given $\kappa$ is (neutrally) stable while the other one is unstable. 

\begin{figure}[t]
\subfigure[\label{fig:l2n1s}]{\includegraphics[width=0.45\textwidth]{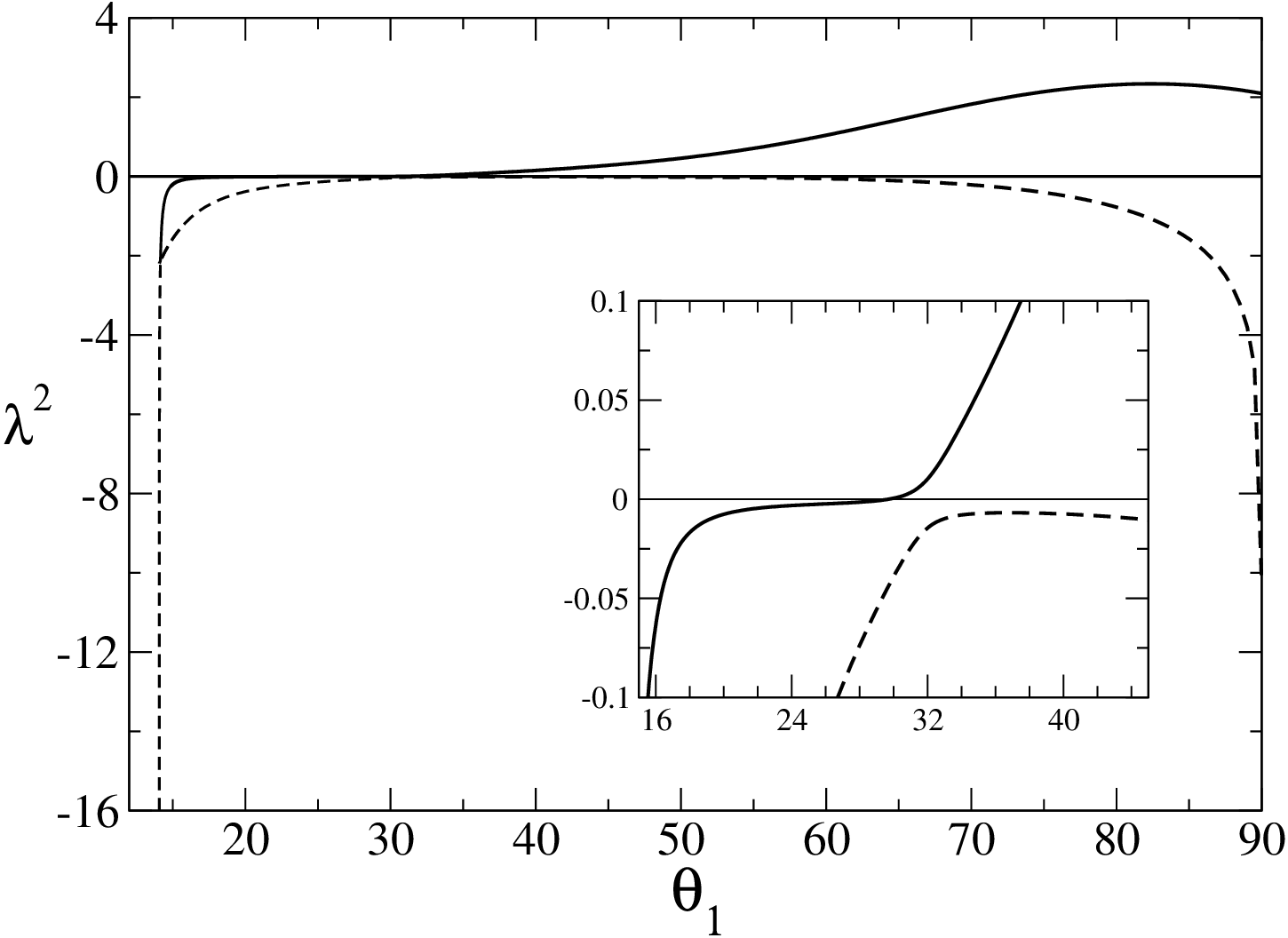}}\qquad
\subfigure[\label{fig:l2n2s}]{\includegraphics[width=0.45\textwidth]{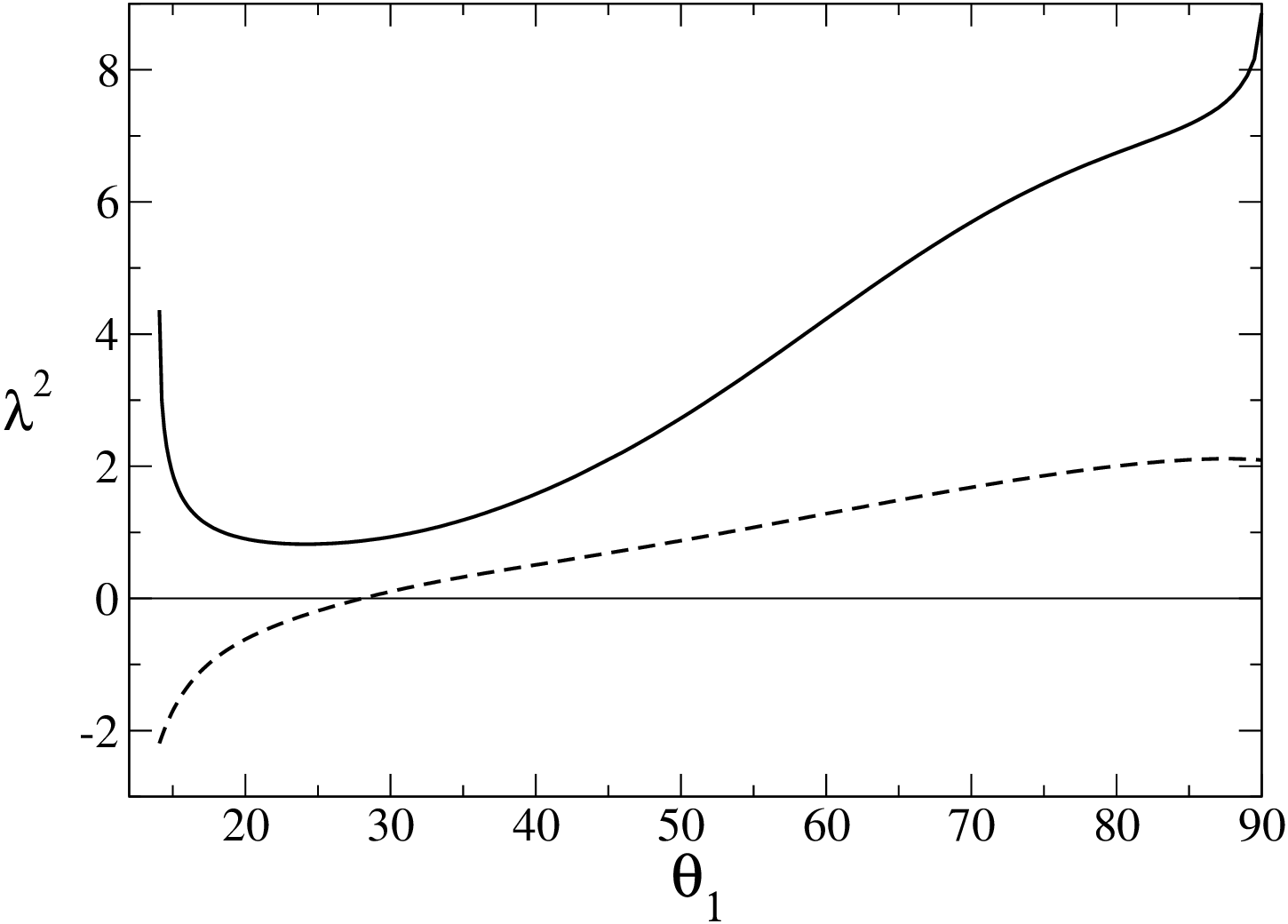}}
\caption{ (a) Squared eigenvalues $\lambda_1^2$ (solid lines) and $\lambda_2^2$ (dashed lines) for symmetric perturbations of the family of asymmetric equilibria indicated by thick solid lines in Fig.~\ref{fig:dk} as a function of the angular position $\theta_1$ of the first vortex. (b) Same for the family of asymmetric equilibria indicated by thin solid lines in Fig.~\ref{fig:dk}.}
\end{figure}

For asymmetric equilibria (i.e., with $\kappa_1\ne\kappa_2$) the expressions for the elements of the matrix $A$ are less manageable than those for the symmetric case and will not be given here. The eigenvalues of $A$ can however be easily computed numerically.  In Fig.~\ref{fig:l2n1s}  we plot the squared eigenvalues $\lambda_1^2$ and $\lambda_2^2$ as function of the angle $\theta_1$ for the family of equilibria indicated by thick solid lines in Fig.~\ref{fig:dk}. From this figure one sees that $\lambda_2^2$ is always negative whereas  $\lambda_1^2$ is negative for $\theta_1<30^\circ$   and positive for $\theta_1>30^\circ$, hence  the equilibria  are neutrally stable in the former region and unstable in the latter. 
In Fig.~\ref{fig:l2n2s} we show the eigenvalues $\lambda_1^2$ and $\lambda_2^2$ as function of the angle $\theta_1$ for the family of equilibria indicated by thin solid lines  in Fig.~\ref{fig:dk}. In this case, $\lambda_1^2$ is positive over the entire range of angles and so these equilibria are always unstable for symmetric perturbations.

\subsubsection{Equilibria on the Normal Line}

\label{sec:4ab}

Computing the eigenvalues of the matrix $A$ for the equilibria on the normal line, one finds that  these equilibria  are of the type saddle-saddle, i.e., $\lambda_1^2>0$ and $\lambda_2^2>0$ (values not shown here). Hence they are unstable under symmetric perturbations. This behavior is analogous to what is observed for a single vortex pair where the corresponding equilibrium on the normal line is a saddle in the symmetric subspace \cite{us}.

\subsubsection{Equilibrium at Infinity}

The  equilibrium at infinity for two vortex pairs  can be viewed as the  combined equilibria of two independent  vortex pairs at $x=\pm\infty$; see Sec.~\ref{sec:infty}. As discussed in detail in Ref.~[4],  the equilibrium at infinity for a single vortex-pair corresponds to a nilpotent saddle, in the sense that the Jacobian matrix $A$ has  two zero eigenvalues with identical eigenvectors.   Similar characterization can be made for the equilibrium at infinity of our four-vortex system, which can thus be referred to as a nilpotent saddle-saddle.

\subsection{Nonlinear Dynamics}
\label{sec:4b}

When restricted to the symmetric subspace, the motion of two pairs of vortices around the cylinder can be described as a Hamiltonian system in a four-dimensional phase space, as already noted.  Although a detailed discussion of this 4D Hamiltonian system is beyond the scope of the present paper, some general observations concerning the  nonlinear dynamics that ensues when the vortices are displaced from their equilibrium positions are in order. Here we shall restrict ourselves to the case of two identical vortex pairs, i.e., $\kappa_1=\kappa_2=\kappa$. We recall that for a given $\kappa$ (in the allowed range) one has two possible equilibrium configurations: a neutrally stable equilibrium (i.e., a center-center) closer behind the cylinder and an unstable equilibrium (either a center-saddle or a saddle-saddle) closer to the cylinder top. 

\begin{figure}[t]
\includegraphics[width=0.45\textwidth]{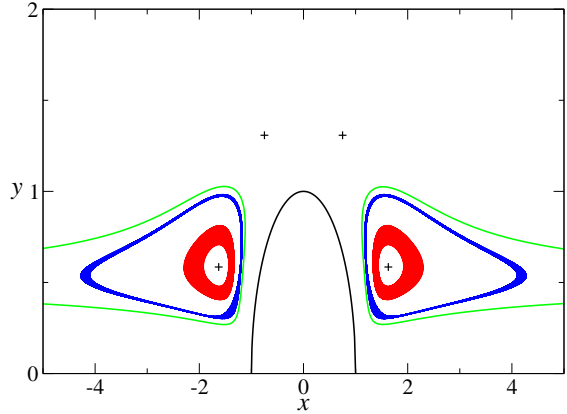}
\caption{(Color online) Vortex trajectories for two pairs of same-signed vortices  with $\kappa=1$ in the symmetric subspace (only the upper vortices are shown).   The initial conditions are: $z_1=2+i0.5$ for the innermost trajectory (mid-gray, red online) and $z_1=4+i0.5$ for the second  trajectory (dark gray, blue online).  
The plus signs indicate the two equilibrium configurations:  $z_1=-\bar{z}_2=1.628+i0.585$ (center-center) and  $z_1=-\bar{z}_2=0.750+i1.308$ (saddle-saddle). The outermost curves (light gray, green online) indicate the nilpotent homoclinic orbit; see text.}
\label{fig:7}
\end{figure}

Small perturbations of the center-center equilibrium will typically result in bounded orbits, as shown
in Fig.~\ref{fig:7} for $\kappa=1$. In this figure, the stable equilibrium is located at $z_1=-\bar{z}_2=1.628+i0.585$ (lower plus signs), the innermost trajectory (mid-gray, red online) corresponds to the initial condition $z_1=-\bar{z}_2=2+i0.5$, and the second trajectory (dark gray, blue online) results from the initial condition  $z_1=-\bar{z}_2=4+i0.5$.  For sufficiently long time each one of these two sets of trajectories tend to fill a compact neighborhood of the equilibrium point. The fact  that these trajectories remain bounded  seems to indicate that this equilibrium may indeed be nonlinearly stable, however further work is necessary to prove nonlinear stability. The outermost curves (light gray, green online) in Fig.~\ref{fig:7} indicate the homoclinic orbit connecting the nilpotent saddle-saddle at infinity. In this case the left and right curves  self-intersect tangentially at the points  $(-\infty,\kappa/2)$ and $(+\infty,\kappa/2)$, respectively. For initial conditions outside this (projected) homoclinic ``loop'' the orbits are generally unbounded. (Some interior initial conditions  may also result in unbounded orbits.)  Note however that each ``orbit''  shown in Fig.~\ref{fig:7}  represent in fact two superimposed plane projections of a 4D orbit, corresponding to the separate trajectories of the two vortices. Thus,  some caution is required when interpreting this and similar figures.

\begin{figure}[t]
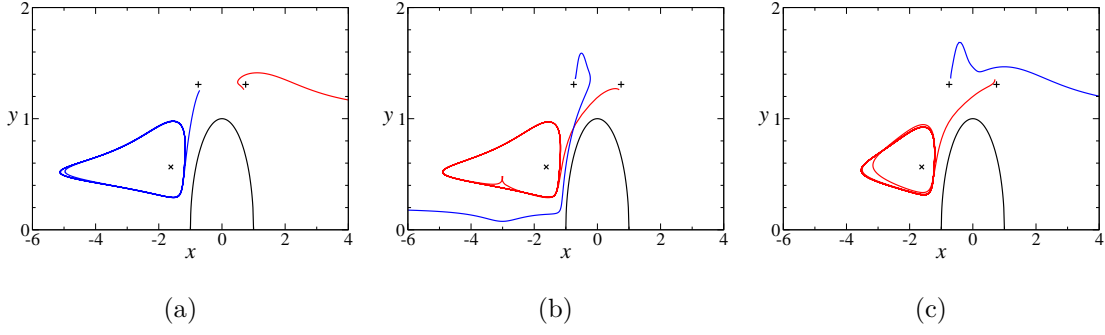

\subfigure[\label{fig:9a}]{\includegraphics[width=0.28\textwidth]{fig12a.eps}}\quad
\subfigure[\label{fig:9b}]{\includegraphics[width=0.28\textwidth]{fig12b.eps}}\quad
\subfigure[\label{fig:9c}]{\includegraphics[width=0.28\textwidth]{fig12c.eps}}
\caption{(Color online)  Vortex trajectories after symmetric perturbations of the saddle-saddle equilibrium (plus signs) for $\kappa=1$. Here  the perturbations  are: (a) $\Delta z_1=-0.05-i0.05$ and $\Delta z_2=0.05-i0.05$, (b) $\Delta z_1=-0.05+i0.05$ and $\Delta z_2=0.05+i0.05$, and (c) $\Delta z_1=-0.05-i0.05$ and $\Delta z_2=0.05+i0.05$.  The cross sign indicates the F\"oppl equilibrium  in front of the cylinder.}
\label{fig:9}
\end{figure}

\begin{figure}[t]
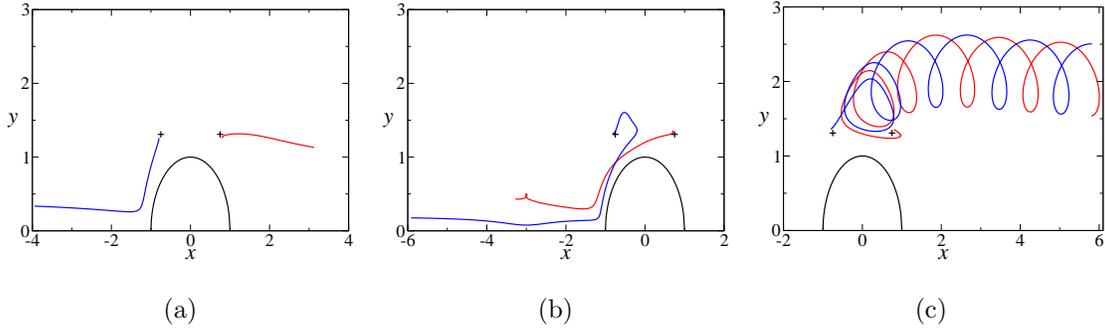

\subfigure[\label{fig:8a}]{\includegraphics[width=0.28\textwidth]{fig13a.eps}}\quad
\subfigure[\label{fig:8b}]{\includegraphics[width=0.28\textwidth]{fig13b.eps}}\quad
\subfigure[\label{fig:8c}]{\includegraphics[width=0.28\textwidth]{fig13c.eps}}
\caption{(Color online) Same as in Fig.~\ref{fig:9} for the following perturbations: (a) $\Delta z_1=0.05-i0.05$ and $\Delta z_2=-0.05-i0.05$, (b) $\Delta z_1=-0.05+i0.05$ and $\Delta z_2=-0.05$, and (c) $\Delta z_1=0.05+i0.05$ and $\Delta z_2=-0.05+i0.05$.}
\label{fig:8}
\end{figure}

Perturbations  of the saddle-saddle equilibrium will, in general, result in unbounded orbits. Here however we have identified two classes of unbounded orbits: i)  semi-bounded orbits, where one of the vortices approaches a periodic orbit  while the other vortex goes to infinity, and ii) completely unbounded orbits, where both vortices go to infinity.  In Fig.~\ref{fig:9} we show examples of semi-bounded orbits. From this figure one sees that one vortex approaches a periodic orbit around the corresponding F\"oppl equilibrium (cross sign) in front of the cylinder, while the other vortex moves to either downstream infinity or upstream infinity depending on the initial condition. (For the situations we have examined, the bounded vortex was never attracted to the F\"oppl equilibrium behind the cylinder, but it is not clear at the moment how general this result is.)  Examples of completely unbounded orbits are shown in Fig.~\ref{fig:8}. Here the vortices can go to infinity either in opposite directions [Fig.~\ref{fig:8a}] or in the same direction. In the latter case, there are two possibilities:  i) the vortices move with constant velocity  with one vortex lagging behind the other [Fig.~\ref{fig:8b}] or ii) they can execute an oscillatory motion  [Fig.~\ref{fig:8c}] whereby  the ``center of mass'' (CM) goes to infinity with constant velocity, while in the CM  reference frame the two vortices trace out an identical circle. Note that this last trajectory is reminiscent of the so-called {\it relative choreographies}  \cite{borisov} performed by ``dancing vortices'' \cite{tokieda} on a plane, in which all vortices follow the same curve when observed from a specific (rotating) reference frame.  Perturbations from the center-saddle will also typically result  in completely unbounded orbits. For certain equilibria in this class it is possible, however, to select specific perturbations that will generate bounded orbits around the co-existing center-center equilibrium, but we shall not pursue this detail here.

\section{Nonsymmetric Dynamics}
\label{sec:4}

\subsection{Antisymmetric Perturbations}

We note that any perturbation of a vortex-pair equilibrium in a two-dimensional flow can be written as the superposition of a symmetric perturbation and an antisymmetric one \cite{marsden,hill1998}. To be precise, consider a generic initial condition for the first vortex pair of the form: $z_1=z_{10}+\Delta z_1$ and $z_3=\overline{z_{10}}+\Delta z_3$, where $z_{10}$ is an equilibrium position and $\Delta z_1$ and $\Delta z_3$ are arbitrary quantities. 
We can  write the perturbations  $\Delta z_1$ and $\Delta z_3$ as
\begin{align}
\Delta z_1 = (\Delta z)_S + (\Delta z)_A \qquad \mbox{and} \qquad \Delta z_3 =\overline{(\Delta z)}_S - \overline{(\Delta z)}_A, 
\label{eq:dz}
\end{align}
where
\begin{align}
(\Delta z)_S=\frac{1}{2}\left(\Delta z_1 +\overline{ \Delta z}_3\right)
\qquad \mbox{and} \qquad
  (\Delta z)_A=\frac{1}{2}\left(\Delta z_1 - \overline{ \Delta z}_3\right).
  \label{eq:sa}
\end{align}
The quantities $(\Delta z)_S$ and $(\Delta z)_A$ correspond to  the symmetric and the antisymmetric components of the perturbation, respectively. Similar expressions hold for the perturbations $\Delta z_2$ and $\Delta z_4$ of the second vortex pair.  In particular, if one considers only antisymmetric perturbations, i.e., $(\Delta z)_S=0$, then Eq.~(\ref{eq:dz}) implies that
\begin{align}
\Delta z_3=-\overline{\Delta z_1}.
%\qquad  \mbox{and}\qquad  \Delta z_4=-\overline{\Delta z_2}.
\label{eq:pas}
\end{align}

 More formally, the decomposition (\ref{eq:dz}) [and respective expressions for the second vortex pair] amounts to saying that the eight-dimensional  tangent space of the phase space of our four-vortex system  can be decomposed \cite{marsden} into a direct sum of a four-dimensional symmetric subspace and its complementary subspace, corresponding to antisymmetric perturbations. Since the symmetric subspace is invariant under the vector field of the linearized system, so is its complementary subspace \cite{marsden}. In other words, the four-dimensional antisymmetric subspace is invariant under the linearized dynamics. Because of this property, we can focus only on the two upper vortices when carrying out the linear stability analysis under antisymmetric perturbations, as discussed next. 

\subsection{Linear Stability Analysis}

Assuming general displacements of the two upper vortices as given in Eq.~(\ref{eq:delta}),  it follows from (\ref{eq:pas}) and similar expression 
%($\Delta z_4=-\overline{\Delta z_2}$) 
for the second vortex pair that the antisymmetric perturbations of the two lower vortices are given by 
\begin{align}
z_3=\bar{z}_{10}-\xi_1+i\eta_1, \qquad z_4=\bar{z}_{20}-\xi_2+i\eta_2.
\label{eq:d2}
\end{align}
Linearizing Eqs.~(\ref{eq:u1}) and (\ref{eq:u2})  for the perturbations given in (\ref{eq:delta}) and (\ref{eq:d2}), one obtains
\begin{equation}
 \left(\begin{array}{c} \dot{\xi_1}\cr \dot{\eta_1}\cr \dot{\xi_2}\cr \dot{\eta_2}\end{array}\right)= 
B  \left(\begin{array}{c}{\xi_1}\cr{\eta_1}\cr{\xi_2}\cr{\eta_2}
\end{array}\right),
\end{equation}
where the matrix $B$ is given by
\begin{equation}
B =\left(
\begin{array}{cccc}
 \frac{\partial u_1}{\partial x_1}-\frac{\partial u_1}{\partial x_3} & \frac{\partial u_1}{\partial y_1}+\frac{\partial u_1}{\partial y_3} & \frac{\partial u_1}{\partial x_2}-\frac{\partial u_1}{\partial x_4} & \frac{\partial u_1}{\partial y_2}+\frac{\partial u_1}{\partial y_4} \\ \\
 \frac{\partial v_1}{\partial x_1}-\frac{\partial v_1}{\partial x_3} & \frac{\partial v_1}{\partial y_1}+\frac{\partial v_1}{\partial y_3} & \frac{\partial v_1}{\partial x_2}-\frac{\partial v_1}{\partial x_4} & \frac{\partial v_1}{\partial y_2}+\frac{\partial v_1}{\partial y_4} \\ \\
 \frac{\partial u_2}{\partial x_1}-\frac{\partial u_2}{\partial x_3} & \frac{\partial u_2}{\partial y_1}+\frac{\partial u_2}{\partial y_3} & \frac{\partial u_2}{\partial x_2}-\frac{\partial u_2}{\partial x_4} & \frac{\partial u_2}{\partial y_2}+\frac{\partial u_2}{\partial y_4} \\ \\
 \frac{\partial v_2}{\partial x_1}-\frac{\partial v_2}{\partial x_3} & \frac{\partial v_2}{\partial y_1}+\frac{\partial v_2}{\partial y_3} & \frac{\partial v_2}{\partial x_2}-\frac{\partial v_2}{\partial x_4} & \frac{\partial v_2}{\partial y_2}+\frac{\partial v_2}{\partial y_4}
\end{array}
\right),
 \end{equation}
 with the derivatives evaluated at the equilibrium positions.  The stability of the respective equilibrium will then depend on the sign of the two squared eigenvalues of the matrix $B$, to be denoted henceforth by $\lambda_3^2$ and $\lambda_4^2$, respectively. In what follows we shall analyze the stability properties of the equilibria described in Sec.~\ref{sec:3aa} with respect to antisymmetric perturbations.

\subsubsection{Same-Signed Equilibria}
\label{sec:5aa}

\begin{figure}[t]
\includegraphics[width=0.5\textwidth]{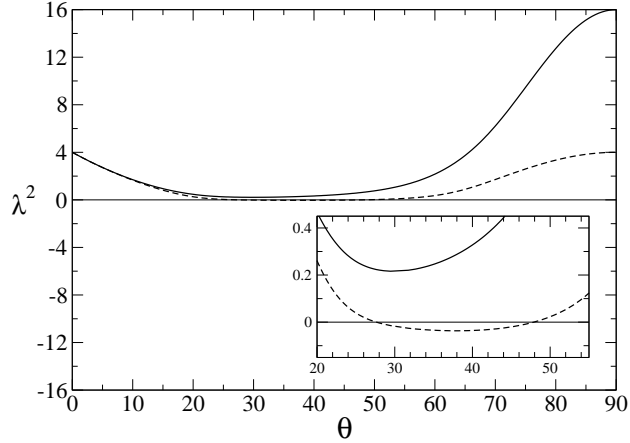}
\caption{Squared eigenvalues $\lambda_3^2$ (solid lines) and $\lambda_4^2$ (dashed lines) for  antisymmetric perturbations for the symmetric as a function of the angle $\theta$ on the curve $C_0$.}
\label{fig:l2a}
\end{figure}

In the case of symmetric  equilibria  it is possible to compute explicitly the elements of  $B$ in terms of rational functions of the equilibrium coordinates $z_0=x+iy$, but the expressions are more cumbersome that those for the matrix $A$ (for symmetric perturbations)  and
will not be presented here. Upon computing the eigenvalues of $B$ along the curve $C_0$ of  symmetric equilibria, one finds the behavior illustrated in Fig.~\ref{fig:l2a}, where we plot the squared eigenvalues $\lambda_3^2$ and $\lambda_4^2$ as a function of the angle $\theta$ on the curve $C_0$. One sees from this figure that $\lambda_3^2$ is always positive, whereas $\lambda_4^2$ is negative in the region  $28^\circ<\theta<48^\circ$ and positive otherwise; see inset of Fig.~\ref{fig:l2a}. In other words, the symmetric  equilibrium is a saddle-saddle  in the antisymmetric subspace for $0<\theta<28^\circ$, bifurcates into a saddle-center for $28^\circ<\theta<48^\circ$, and reverts to a saddle-saddle for $48^\circ<\theta<90^\circ$.  This implies  that the symmetric equilibrium is always unstable against antisymmetric perturbations. 
 
\begin{figure}[t]
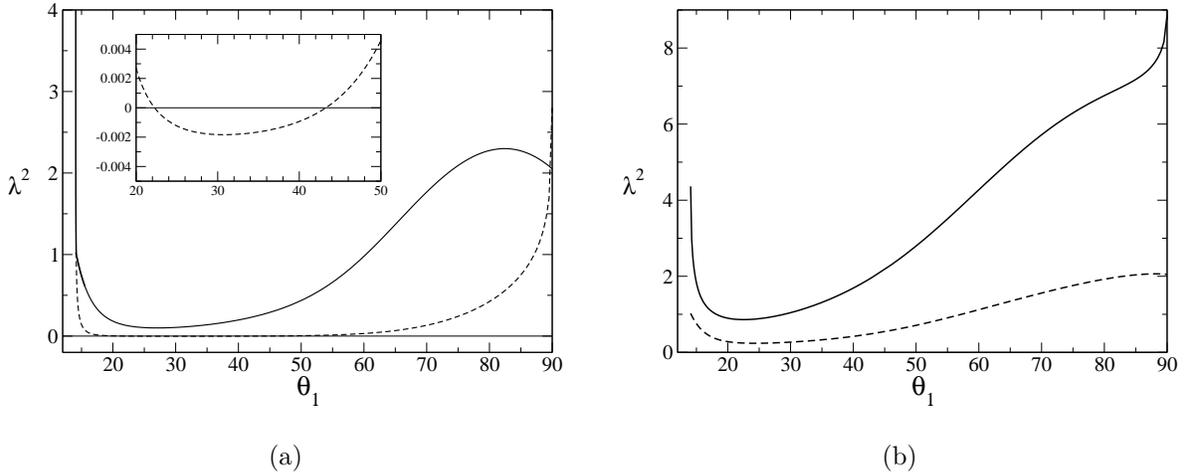

\vspace{0.3cm}
\subfigure[\label{fig:l2n1a}]{\includegraphics[width=0.45\textwidth]{fig15.eps}}\qquad
\subfigure[\label{fig:l2n2a}]{\includegraphics[width=0.45\textwidth]{fig16.eps}}
\caption{ (a) Squared eigenvalues $\lambda_3^2$ (solid lines) and $\lambda_4^2$ (dashed lines) for antisymmetric perturbations of the  family of asymmetric equilibria indicated by thick solid lines in Fig.~\ref{fig:dk} as a function of the angular position $\theta_1$ of the first vortex. (b) Same for the second family of asymmetric equilibria (thin solid lines in Fig.~\ref{fig:dk}).}
\end{figure}

In Fig.~\ref{fig:l2n1a} we show the squared eigenvalues $\lambda_3^2$ and $\lambda_4^2$ as function of the angle $\theta_1$ for the  family of asymmetric equilibria indicated by thick solid lines in Fig.~\ref{fig:dk}.
Here the behavior is similar to what is found for the symmetric equilibria (compare with Fig.~\ref{fig:l2a}), in the sense that $\lambda_3^2$ is always positive while $\lambda_4^2$ becomes slightly negative in a small region of angles [see inset of Fig.~\ref{fig:l2n1a}], and so the equilibrium is unstable in the antisymmetric subspace.   
In Fig.~\ref{fig:l2n2a} we plot the squared eigenvalues $\lambda_3^2$ and $\lambda_4^2$  for the family of asymmetric equilibria  indicated by thin solid lines in Fig.~\ref{fig:dk}. Here both squared eigenvalues are positive and hence these equilibria are also unstable for antisymmetric perturbations.

 \subsubsection{Equilibria on the Normal Line}

As in the case of symmetric perturbations discussed in Sec.~\ref{sec:4ab}, the squared eigenvalues  of the matrix $B$ for the equilibria on the normal line  are positive, implying a  saddle-saddle equilibrium in the antisymmetric subspace. Thus, the equilibrium  for two vortex pairs on the normal line is unstable with respect to both symmetric and antisymmetric perturbations. This behavior is reminiscent of the fact \cite{us} that the equilibrium on the normal line for a single vortex pair is unstable (i.e., a saddle) for both types of perturbations.

\section{Discussions}
\label{sec:5}

\begin{figure}
\includegraphics[width=0.45\textwidth]{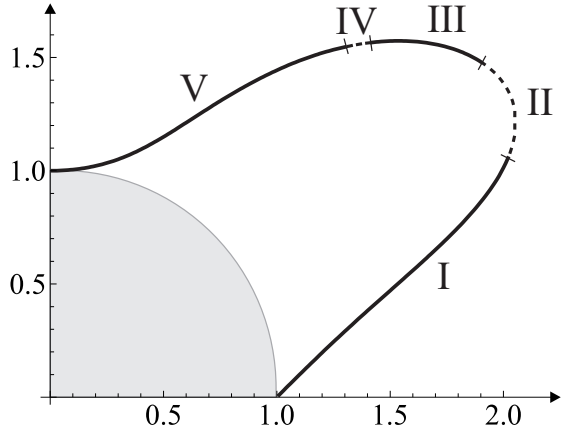}
\caption{Regions of different stability properties of the symmetric equilibria; see Table I for a description of the five regions indicated in the figure.}
\label{fig:statp4v}
\end{figure}

Here we wish to discuss in more details the  stability properties of the same-signed equilibria of our four vortex system and comment on the possible practical implications of our results.
First we consider the symmetric equilibria for which the two vortex pairs have  equal strength.
Let us also recall that  $\lambda_1^2$ and $\lambda_2^2$ refer to the two pairs of eigenvalues for the symmetric modes, whereas  $\lambda_3^2$ and $\lambda_4^2$ correspond to the  eigenvalues for the antisymmetric modes.  

In view of the changes of sign of $\lambda_i^2$  along the locus $C_0$ of symmetric  equilibria, see Figs.~\ref{fig:l2s} and \ref{fig:l2a}, it is convenient to divide this curve into five different regions according to the stability properties of the equilibrium, as shown in Fig.~\ref{fig:statp4v}. The nature of the equilibrium in each of these regions is indicated in Table~\ref{tab}, where we denote with the letter C (from ``center'') the regions where  $\lambda_i^2<0$, implying a pair of purely imaginary eigenvalues, and with the label S (from ``saddle'') the regions for which $\lambda_i^2>0$, giving a pair of real eigenvalues of opposite signs.

\begin{table}[b]
\begin{tabular}{|c||c|c|c|c|c|}
%\tabhead{6}{7.6cm} %%required
\hline
Region & I & II & III & IV & V \\
\hline
Interval & $0<\theta<28^\circ$ & $28^\circ<\theta<38^\circ$ & $38^\circ<\theta<48^\circ$ & $48^\circ<\theta<50^\circ$ & $50^\circ<\theta<90^\circ$ \\
\hline
$\lambda_1^2$ & C & C & S & S & S \\
\hline
$\lambda_2^2$ & C & C & C & C & S \\
\hline
$\lambda_3^2$ & S & S & S & S & S \\
\hline
$\lambda_4^2$ & S & C & C & S & S \\
\hline 
\end{tabular}
\caption{Regions of stability along the locus $C_0$ of the symmetric equilibria. Here C denotes ``center'' ($\lambda_i^2<0$) and S denotes ``saddle'' ($\lambda_i^2>0$). The angle intervals are approximation within two significant digits.}
\label{tab}
\end{table}

From a practical standpoint, the most relevant information in Table I is perhaps the fact that the equilibrium configurations  located in regions I and II are neutrally stable under symmetric perturbations.
Since symmetry  can be enforced by placing a splitter plate in the middle plane behind the cylinder  \cite{us}, this equilibrium could in principle be observed in experiments. Of course, the difficulty here is to generate vortices in front of the cylinder. This may however be possible by placing a sufficiently long splitter plate in front of the cylinder, which would have the tendency of generating vortices in front of the cylinder with the same sign of the vortices behind it. Now, even if one manages to produce vortex pairs in front and behind the cylinder, it is unlikely that they would have  the  same strength. In this context, it is interesting to note that there is a subset of asymmetric equilibria that are stable in the symmetric subspace, and those could in principle have a physical counterpart. It nonetheless remains an open question whether four-vortex  configurations with one pair of recirculating eddies on each side of the cylinder can be observed in real flows.

The question of the possible existence of equilibria for two opposite-signed vortex pairs on the opposite sides of the cylinder is also of practical interest.  We have seen in  Sec.~\ref{sec:normal}  that there are no  such equilibria in an unbounded plane.  This means, in particular, that the four-vortex configuration observed in the counterflow of superfluid helium past a cylinder  \cite{nature} cannot be realized in the flow of a classical fluid past a cylinder in an otherwise unbounded region. Such configurations might however exist for flows past a cylinder confined in a rectangular channel, which was in fact the geometry used in the experiments \cite{nature}. In this case, the channel walls tend to generate vortices in front of the cylinder with the opposite sign of the vortices behind it. Indeed, recirculating eddies in front of a circular cylinder placed near a plane boundary have been observed \cite{lin} when the gap between the cylinder and the plane is sufficiently small.  It thus seems possible that confining the cylinder between two plane walls may induce the formation of  a vortex pair  in front of the cylinder with the opposite polarity of the pair behind it.

From a theoretical perspective, the treatment of point-vortex dynamics in the presence of a cylinder placed between two plane walls is a much more complicated problem because of the infinitely many vortex images that one has to consider. We are currently investigating this problem. The existence of stationary configurations for opposite-signed vortex pairs on the opposite sides of the cylinder (if found) would be of considerable interest because it could explain the four-vortex configurations  reported  in Ref.~[11] entirely within the scope of classical fluid mechanics, without having to invoke the two-fluid model of superfluid helium. 

\section{Conclusions}
\label{sec:6}

We have presented a detailed study of the stationary configurations and their stability for two pairs of point vortices  placed in a uniform flow past a circular cylinder. We have shown that among the possible same-signed equilibria there exists a large subset of configurations that, although unstable under generic perturbations, are stable with respect to symmetric perturbations. The nonlinear dynamics within the symmetric subspace was also studied, and here we found three general classes of orbits: i) bounded orbits around the stable equilibrium, ii) semi-bounded orbits where one of the vortex pairs is attracted to the F\"oppl equilibrium while the other pair goes to infinity, and iii) completely unbounded orbits where both vortex pairs move to infinity. We have obtained, furthermore, a previously unknown set of opposite-signed equilibria  for which the vortices lie on the line bisecting the cylinder perpendicularly to the incoming flow. 
Finally, we have suggested  that if the cylinder  is  confined in a rectangular channel  then  equilibrium configurations for two opposite-signed vortex pairs should exist  with one  vortex pair windward  of the cylinder and the other pair in the leeward side. The existence of such  equilibrium could explain the unusual four-vortex configuration recently observed  for the counterflow of superfluid helium past a cylinder.

\section{Acknowledgements}
This work was supported financially  in part by the Brazilian agencies CNPq and FACEPE.

\appendix

\section{The Locus of Same-Signed Equilibria}
\label{app:A}

%\begin{widetext}

After setting $\kappa_1=\kappa_2=\kappa$ in Eq.~(\ref{eq:u1}), taking the real and imaginary parts, and solving for $u_1=0$ and $v_1=0$, one finds after some manipulation that the equilibrium positions for the first vortex lie on the curve given by $P(x,y)=0$, where
\begin{align}
P(x,y)&=\left(2+y^2\right) r^{12}+\left(2-16 y^2\right) r^{10}+ \left(-4-19 y^2+16 y^4\right)r^{8}\cr &+4  \left(-1-8 y^2+8 y^4\right)r^{6}+ \left(2+19 y^2+96 y^4-16 y^6\right)r^4 \cr & -2  \left(-1+8 y^2+16 y^4+32 y^6\right)r^2+ y^2 \left(-1+16 y^2+16 y^4\right),
\label{eq:P2}
\end{align}
with $r^2=x^2+y^2$.
The corresponding  vortex intensity is given by
\begin{equation}
\kappa=\frac{Q(x,y)}{R(x,y)},
\end{equation}
where the polynomials $Q(x,y)$ and $R(x,y)$ take the form 
%\begin{widetext}
\begin{align}
Q(x,y)&=2 y \left(-1+x^4+2 x^2 y^2+y^4\right) \cr &\times \left(-x^6+\left(-1+y^2\right) \left(1+y^2\right)^2 +x^4 \left(1+15 y^2\right) +x^2 \left(1-6 y^2-15 y^4\right)\right),
\label{eq:Q}
\end{align}
%\end{widetext}
and
%\begin{widetext}
\begin{align}
R(x,y)&=
x^{10}-12 x^8 y^2+4 x^4 y^2 \left(3+10 y^2\right)-2 x^6 \left(1+4 y^2+11 y^4\right)\cr & +4 y^4 \left(2-y^2-2 y^4+y^6\right)+x^2 \left(1-8 y^2-6 y^4+40 y^6+13 y^8\right).
\label{eq:R}
\end{align}
%\end{widetext}

\section{The Matrix $A$ for Symmetric Equilibria}
\label{app:B}

The matrix $A$ calculated at an equilibrium point $(x,y)$ for two identical vortex pairs has the following six independent elements:
\begin{align}
A_{11}=&\frac{2 \left(x^3-3 x y^2\right)}{\left(x^2+y^2\right)^3}+\frac{\kappa}{4}  \left[\frac{x}{x^2+(-1+y)^2}+\frac{2 x (-1+y)}{\left(x^2+(-1+y)^2\right)^2}-\frac{4 (-1+x) y}{\left((-1+x)^2+y^2\right)^2}\right.\cr 
&+ \frac{8 x y}{\left(-1+x^2+y^2\right)^2}+\frac{2 x y}{\left(x^2+y^2\right)^2}-\frac{8 x y}{\left(1+x^2+y^2\right)^2}-\frac{4 (1+x) y}{\left((1+x)^2+y^2\right)^2}\cr 
&+ \left.\frac{2 x (1+y)}{\left(x^2+(1+y)^2\right)^2}-\frac{x}{x^2+(1+y)^2}\right],
\end{align}
\begin{align}
A_{12}=&\frac{2 \left(3 x^2 y-y^3\right)}{\left(x^2+y^2\right)^3}+\frac{\kappa}{4} \left[\frac{1}{x^2}-\frac{2 x^2}{\left(x^2+(-1+y)^2\right)^2}+\frac{2}{y^2}+\frac{y}{x^2+(-1+y)^2}\right.\cr 
&+ \frac{4 (-1+x)^2}{\left((-1+x)^2+y^2\right)^2}-\frac{2}{(-1+x)^2+y^2}-\frac{8 \left(-1+x^2\right)}{\left(-1+x^2+y^2\right)^2}+\frac{4}{-1+x^2+y^2}\cr 
&+ -\frac{2 x^2}{\left(x^2+y^2\right)^2}+\frac{1}{x^2+y^2}+\frac{4+8 x^2}{\left(1+x^2+y^2\right)^2}-\frac{4}{1+x^2+y^2}+\frac{4 (1+x)^2}{\left((1+x)^2+y^2\right)^2}\cr 
& \left.-\frac{2}{(1+x)^2+y^2}-\frac{2 x^2}{\left(x^2+(1+y)^2\right)^2}-\frac{y}{x^2+(1+y)^2}\right],
\end{align}
\begin{align}
A_{13}=-\frac{\kappa x y \left(-1+x^4+y^4+2 x^2 \left(-2+y^2\right)\right) \left(-1+4 y^2+\left(x^2+y^2\right)^2\right)}{2 \left(x^2+(-1+y)^2\right)^2 \left(x^2+y^2\right)^2 \left(x^2+(1+y)^2\right)^2}, 
\end{align}
\begin{align}
A_{14}=&\frac{\kappa}{4}  \left[-\frac{1}{x^2}+\frac{2-y}{x^2+(-1+y)^2}+\frac{1}{x^2+y^2}+\frac{4}{\left(1+x^2+y^2\right)^2}+\frac{2+y}{x^2+(1+y)^2}\right.\cr 
&-\left. 2x^2 \left(\frac{1}{\left(x^2+(-1+y)^2\right)^2}+\frac{1}{\left(x^2+y^2\right)^2}+\frac{1}{\left(x^2+(1+y)^2\right)^2}\right)\right],
\end{align}
\begin{align}
A_{21}=&\frac{2 \left(3 x^2 y-y^3\right)}{\left(x^2+y^2\right)^3}+\frac{\kappa}{4} \left[ \frac{1}{x^2}-\frac{2 x^2}{\left(x^2+(-1+y)^2\right)^2}+\frac{y}{x^2+(-1+y)^2}+\frac{4 (-1+x)^2}{\left((-1+x)^2+y^2\right)^2}\right.\cr
&-\frac{2}{(-1+x)^2+y^2}-\frac{8 x^2}{\left(-1+x^2+y^2\right)^2}+\frac{4}{-1+x^2+y^2}-\frac{2 x^2}{\left(x^2+y^2\right)^2}+\frac{1}{x^2+y^2}
\cr
&+\frac{4+8 x^2}{\left(1+x^2+y^2\right)^2}-\frac{4}{1+x^2+y^2}+\frac{4 (1+x)^2}{\left((1+x)^2+y^2\right)^2}-\frac{2}{(1+x)^2+y^2}\cr
&-\left.\frac{2 x^2}{\left(x^2+(1+y)^2\right)^2}-\frac{y}{x^2+(1+y)^2}\right],
\end{align}
\begin{align}
A_{23}=&\frac{\kappa}{4} \left[-\frac{1}{x^2}+\frac{-2+y}{x^2+(-1+y)^2}-\frac{1}{x^2+y^2}+\frac{4}{\left(1+x^2+y^2\right)^2}-\frac{2+y}{x^2+(1+y)^2}\right. \cr
&+\left.2 x^2 \left(\frac{1}{\left(x^2+(-1+y)^2\right)^2}+\frac{1}{\left(x^2+y^2\right)^2}+\frac{1}{\left(x^2+(1+y)^2\right)^2}\right)\right].
\end{align}

The remaining elements are obtained from the following relations:
\begin{equation}
A_{22}=-A_{11}, \qquad
A_{24}=A_{13},
\end{equation}
\begin{equation}
A_{31}=-A_{13}, \qquad A_{32}=A_{14}, \qquad A_{33}=-A_{11}, \qquad
A_{34}=A_{12},
\end{equation}
\begin{equation}
A_{41}=A_{23}, \qquad A_{42}=-A_{13}, \qquad A_{43}=A_{21}, \qquad A_{44}=A_{11}.
\end{equation}

\end{document}